# In silico study on the cytotoxicity against Hela cancer cells of xanthones bioactive compounds from *Garcinia cowa*: Network Pharmacology, QSAR based on Graph Deep Learning, and Molecular Docking

Nguyen Manh Son[1], Pham Huu Vang[1],
Nguyen Thi Dung[1], Nguyen Manh Ha[1], Ta Thi Thao[1], Tran Thi Thu Thuy[2], Phan Minh Giang[1]

[1]Hanoi University of Science, Vietnam National University, Hanoi, Vietnam
[2]Institute of Natural Products Chemistry, Vietnam Academy of Science and Technology, 18 Hoang Quoc Viet, Nighiado, Cau Giay, Hanoi, Vietnam

**Abstract:** Cancer is recognized as a complex group of diseases, contributing to the highest global mortality rates, with increasing prevalence and a trend toward affecting younger populations. It is characterized by uncontrolled proliferation of abnormal cells, invasion of adjacent tissues, and metastasis to distant organs. Garcinia cowa, a traditional medicinal plant widely used in Southeast Asia, including Vietnam, is employed to treat fever, cough, indigestion, as a laxative, and for parasitic diseases. Numerous xanthone compounds isolated from this species exhibit a broad spectrum of biological activities, with some showing promise as anti-cancer and antimalarial agents. Network pharmacology analysis successfully identified key bioactive compounds Rubraxanthone, Garcinone D, Norcowanin, Cowanol, and Cowaxanthone—alongside their primary protein targets (TNF, CTNNB1, SRC, NFKB1, and MTOR), providing critical insights into the molecular mechanisms underlying their anti-cancer effects. The Graph Attention Network algorithm demonstrated superior predictive performance, achieving an $R^2$ of 0.98 and an RMSE of 0.02 after data augmentation, highlighting its accuracy in predicting pIC50 values for xanthone-based compounds. Additionally, molecular docking revealed MTOR as a potential target for inducing cytotoxicity in HeLa cancer cells from Garcinia cowa.

## I. Introduction

Cancer is a complex group of diseases and one of the leading causes of mortality worldwide, characterized by the uncontrolled proliferation of abnormal cells, the ability to invade adjacent tissues, and metastasis to distant organs in the body [1, 2]. The development of cancer involves multiple molecular and genetic disruptions, leading to imbalances in signaling pathways that regulate the cell cycle, programmed cell death (apoptosis), and DNA repair [3]. In this context, cervical cancer is one of the most prevalent cancers among women, particularly in developing countries, and is a significant focus of biomedical research [4]. The HeLa cell line, isolated from a malignant cervical tumor of patient Henrietta Lacks in 1951, has become one of the most critical in vitro models in cancer research [5]. HeLa cells are distinguished by their infinite proliferative capacity, a trait that enables continuous division in culture conditions without undergoing typical cellular senescence. Furthermore, their relative genetic stability and adaptability to experimental conditions have made HeLa cells an ideal tool for studying the molecular biology of cancer, from gene regulation to signaling pathways involved in cancer cell survival and proliferation [6].

The Clusiaceae family (formerly Guttiferea) consists of woody plants thriving in wet forests, often near water streams. It includes about 800 tree species distributed worldwide, primarily in tropical regions [7]. The family is divided into three tribes Clusiaceae, Garcinia, and Symphonies and is easily identified and classified. Garcinia cowa typically grows upright with numerous branches. Its leaves are obovate, measuring 7-17 cm long and 2.5-6 cm wide. Petioles are about 2 cm long and slender. The male inflorescence consists of 3-8 flowers arranged in a canopy, with 1 cm long peduncles, 4 sepals, 4 thick petals, and stamens clustered in a block. The fruit is usually stout, spherical, slightly flattened, with visible citrus-like segments [8]. This tropical fruit tree grows wild along forest edges in Southeast Asia. In Vietnam, it is found in the mountainous forests of the northern and central regions. Garcinia cowa is a traditional medicinal plant widely used in Southeast Asia, including Vietnam, to treat fever, cough, indigestion, as a laxative, and for parasitic diseases. Many xanthone compounds have been isolated from this species, exhibiting a broad spectrum of biological activities, with several showing potential as anti-cancer and antimalarial agents [9].

Network pharmacology is an advanced scientific field that integrates systems biology theory with network analysis tools, bioinformatics, and traditional pharmacology to study the complex interactions between drugs and biological targets from a holistic systems and network biology perspective [10, 11]. This field enables researchers to explore multidimensional molecular relationships, including drug-target, drug-drug, and drug-biological system interactions, providing deep insights into the mechanisms of drug action at a system-wide level [12, 13]. In recent years, network pharmacology has become a powerful tool in uncovering systemic pharmacological mechanisms, supporting the development of new therapeutic strategies, optimizing clinical therapies, and accelerating drug discovery [14, 15]. The rapid advancement of artificial intelligence has significantly enhanced the potential of network pharmacology [16]. With powerful computational capabilities and deep learning algorithms, AI has overcome the inherent limitations of traditional methods, which are often inefficient and less accurate in screening and identifying potential drug molecules [17, 18]. Advanced AI models can analyze vast amounts of biological data, predict drug-target interactions, assess toxicity, and propose new compounds with higher precision [19]. Moreover, AI can integrate multi-omics data, enabling the construction of more detailed biological network models, thereby improving the prediction of drug efficacy and safety [20, 21]. Additionally, molecular docking involves predicting the interactions between a small molecule and a protein at the atomic level [22, 23]. This allows researchers to study the behavior of small molecules, such as nutrients, in the binding site of a target protein and understand the fundamental biochemical processes underlying these interactions [24].

## II. Materials and methods

### 2.1. Hela anti-cancer activities experiment results

| **Compounds** | **IC50 (μM)** |
|---|---|
| Fuscaxanthone A | No data |
| 7-O-methyl garcinone E | No data |
| Cowagarcinone A | No data |
| Cowanin | 16.58 |
| Cowaxanthone | 15.75 |
| Rubraxanthone | 7.85 |
| Norcowanin | 13.46 |
| Cowanol | 12.19 |
| Garcinone D | 11.96 |
| Fuscaxathone I | 45.86 |
| Garcicowanone F | 46.55 |
| Garcicowanone G | > 50 |
| Garcicowanone H | > 50 |
| Kaennacowanol A | > 50 |
| Garcinone F | 19.28 |
| Norcowanone A | > 50 |
| Norcowanone B | > 50 |
| Garcicowanone C | > 50 |
| Garcicowanone D | > 50 |
| Garcicowanone E | > 50 |
| Garcicowanone I | > 50 |
| α-mangostin | No data |

Table 1: Experimental results of cytotoxicity activity on HeLa cells

## 2.2. Network Pharmacology

### 2.2.1. Targets prediction of active compounds of Garcinia cowa

The active ingredients from Garcinia cowa have been constructed through synthesis based on experimental analysis and research from scientific publications. The search for scientific publications involved selecting database sources from ScienceDirect, and Google Scholar using the keyword "Garcinia cowa". The documents were then carefully filtered to build a dataset of active components. The predicted drug targets were based on the SwissTargetPrediction tool.

### 2.2.2. Acquisition of potential therapeutic targets

In order to identify potential therapeutic targets, the keyword ”hela” was used to search through multiple databases, including the Genecard database. This process involved collecting and screening relevant targets associated with ’hela’ to ensure comprehensive coverage of possible therapeutic options.

### 2.2.3. Acquisition of Garcinia cowa targets in Hela cancer cells

To clarify the mechanism of action of drug targets and disease targets at the protein level, the drug's gene targets and disease-related genes were incorporated into a Venn diagram using an open-source tool developed in Python programming language. This tool helps identify common targets, which are the intersections of disease-related targets and drug-related targets. The active compounds with the most targets will be assumed to be the primary bioactive compounds responsible for the activity in the medicinal plant.

### 2.2.4. Protein-protein interaction network

Protein-protein interaction networks help to better understand the biological mechanisms related to disease pathogenesis, specifically focusing on target-pathogenesis interactions at the protein level. The STRING database was used to build the data for the protein-protein interaction network by providing common targets identified through the Venn diagram. The interaction network data was then visualized and analyzed using an open-source tool developed in Python programming language. The main targets were determined based on the importance of the nodes, including Degree Centrality, Betweenness Centrality, Closeness Centrality, PageRank, and Composite Score. The higher these values, the greater the role of the gene in the interaction network. Next, proteins with the highest scores were selected as the next targets for further research.

### 2.2.5. GO and KEGG analysis

The identified main targets were analyzed using GO-BP (Gene Ontology - Biological Process), GO-CC (Gene Ontology - Cellular Component), GO-MF (Gene Ontology - Molecular Function), and KEGG pathways through ShinyGO version 0.76 (http://bioinformatics.sdstate.edu/go76/).

## 2.3. Machine learning for predicting cytotoxicity

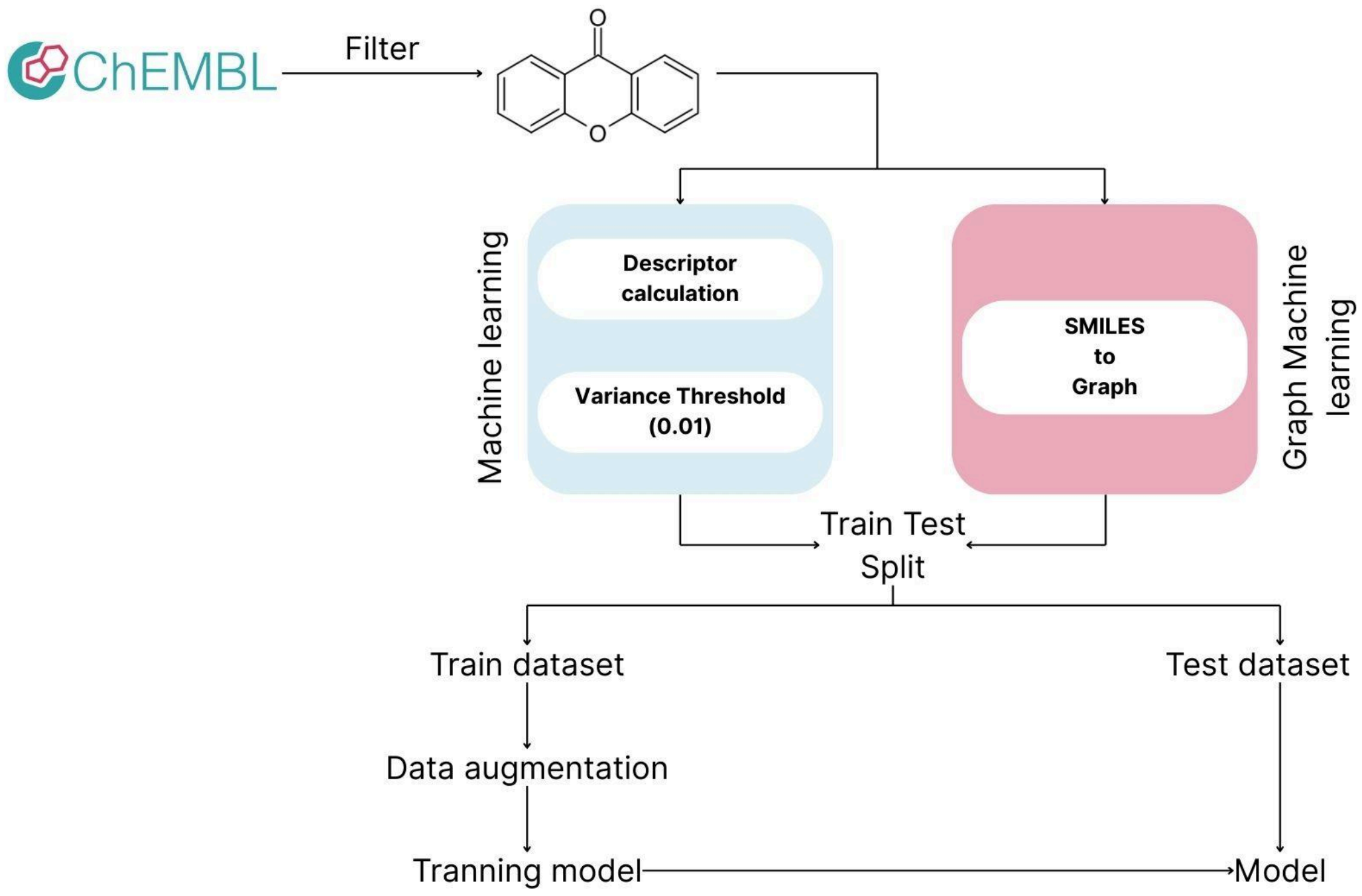

Figure 1: Quy trình làm việc của mô hình học máy truyền thống và học máy đồ thị

### 2.3.1. Data collection and preprocessing

The bioactivity data were collected from the ChEMBL database (www.ebi.ac.uk/chembl/), a public data source containing information on chemical compounds and their biological activities. To collect the data, we used the keyword "HeLa" (a cervical cancer cell line) to filter for compounds associated with this biological target, focusing on activity measurements such as IC50 values. The datasets were downloaded in raw format and underwent rigorous preprocessing to ensure quality and suitability for subsequent analysis steps. This included removing duplicate entries, eliminating SMILES strings longer than 75 characters to reduce computational complexity and focus on moderately sized molecules, removing compounds containing salts or ionized forms, and filtering out invalid SMILES strings. After preprocessing, valid SMILES strings were used to calculate molecular descriptors using

RDKit (www.rdkit.org/), which served as input features for machine learning models. To reduce data dimensionality and remove non-informative features, the Variance Threshold method was applied. This technique eliminates features with low variance, helping to improve computational efficiency and prevent overfitting in machine learning models. Following the train-test split, the training dataset was augmented using the Bootstrapping method for traditional machine learning models, and through random generation of new, highly similar molecules along with noise addition for graph-based machine learning models.

### 2.3.2. Decision Tree

A decision tree is a machine learning model that represents decisions in a tree structure, where each node represents a condition based on a feature, each branch denotes the outcome of the condition, and each leaf represents a class label or predicted value [25, 26, 27]. The algorithm selects the optimal feature and threshold at each node to split the data, typically based on criteria such as Gini impurity or entropy, to minimize heterogeneity in the resulting subsets [28]. The final tree consists of decision nodes and leaf nodes. A decision node has two or more branches, each representing values for the feature being tested, while a leaf node indicates a decision about the target. The topmost decision node is called the root node.

Suppose that at node $m$, there is a set represented by $Q_m$, consisting of $N_m$ samples. For each data split θθ $= (j, t_m)$, which includes feature $j$ and splitting threshold $t_m$, the data is divided into two subsets:

$$Q_m^{left}(\theta) = \{x = \{x_1, x_2, ..., x_i, ..., x_n\} \in Q_m \mid x_i \le t_m\}$$

$$Q_m^{right}(\theta) = Q_m \setminus Q_m^{left}(\theta)$$

The quality of the split is then evaluated using one of two criteria: Gini impurity or entropy (for classification tasks) and Mean Squared Error (MSE) for regression tasks [29]. Let $N_{m,c}$ denote the number of elements belonging to class $c$ in the set $Q_m$.

The Gini impurity criterion is defined as follows:

$$Gini = 1 - \sum_{c=1}^{C} \left(\frac{N_{m,c}}{N_m}\right)^2$$

The Entropy criterion is defined as follows:

$$Entropy = - \sum_{c=1}^{C} \frac{N_{m,c}}{N_m} log_2 \frac{N_{m,c}}{N_m}$$

The process of building the decision tree continues until stopping conditions are met, such as reaching a maximum depth, a minimum number of samples at a node, or no significant improvement in the split. Although decision trees are interpretable and intuitive, they are prone to overfitting if not properly tuned. To prevent overfitting and ensure the model maintains generalization ability, stopping criteria are carefully applied. These criteria control the complexity of the tree by setting limits on the minimum number of data points in a node or leaf before splitting, while also restricting the depth of the tree [30, 31, 32].

### 2.3.3. Random Forest

Random Forest is an ensemble learning technique based on decision trees, utilizing the bagging method [33]. In this approach, a collection of decision trees is trained on random subsets of the original dataset, sampled with replacement. Additionally, at each split node of every tree, only a random subset of features is considered, which enhances randomness and reduces correlation between trees. The prediction process is performed by aggregating the outputs of the trees, using averaging for regression tasks or majority voting for classification tasks. The prediction process aggregates the outputs of all trees. For regression tasks, the final prediction is computed as the average of individual tree predictions, expressed mathematically as:

$$y = \frac{1}{T}\sum_{t=1}^{T} h_t(x)$$

where $h_t(x)$ is the predicted value of the t-th tree for the input x, and T is the total number of trees.

The Random Forest model generates predictions by averaging the outputs of multiple decision trees, each trained on either the entire training dataset or different subsets of it and/or on either all features or different subsets of features. This approach, similar to bagging techniques, helps reduce variance, thereby mitigating the overfitting problem commonly observed in individual decision trees [29]. Overall, the final model achieves higher performance by accepting a slight increase in bias and sacrificing some interpretability. The training and prediction processes of Random Forest are fast due to the simplicity of the base decision trees and can be easily parallelized because of the complete independence between the trees.

**2.3.4. XGBoost**

XGBoost (Extreme Gradient Boosting) is an optimized version of the Gradient Boosting algorithm, designed to enhance efficiency and accuracy during model training [34] .The algorithm constructs the model iteratively, where each new tree is trained to correct the errors of previous trees using gradient descent [35]. At the t-th iteration, the model is updated as follows:

$$\hat{y}_i^{(t)} = \hat{y}_i^{(t-1)} + f_t(x_i), f_t \in F$$

Where $F$ is the space of regression trees.

The overall objective function optimized at step $t$ is given by:

$$L^{(t)} = \sum_{i=1}^{n} l(y_i, \hat{y}_i^{(t)}) + \sum_{k=1}^{t} \Omega(f_k)$$

Here $\Omega(f_k) = \gamma T + \frac{1}{2}\lambda \sum_{j=1}^{T} w_j^2$ is the regularization term, comprising the number of leaves $T$ and the weights $w$ of each leaf.

Leveraging second-order derivative approximation (Taylor expansion), XGBoost achieves more efficient optimization compared to traditional boosting methods. Additionally, features such as handling missing values, parallel computation, reverse tree pruning, and early stopping enhance the model's practical performance.

**2.3.5. Graph Neural Network**

Graph Neural Networks (GNNs) are a class of deep learning models designed to process graph-structured data, where nodes represent entities and edges denote relationships between them [36]. GNNs operate based on a message-passing mechanism, where each node's features are updated by aggregating information from its neighboring nodes across multiple layers [37]. Specifically, at layer $k + 1$ the feature vector of node v v v is computed as follows:

$$h_v^{(k+1)} = \sigma(W^{(k)} \cdot AGGREGATE(\{h_u^{(k)}, \forall u \in N(v) \cup \{v\}\}) + B^{(k)} h_v^{(k)})$$

$$\mathcal{N}(v)$$

Where $h_v^{(k)}$ is the feature vector of node $v$ at layer $k$, is the set of neighboring nodes, AGGREGATE is an aggregation function (typically sum, mean, or max), $W^{(k)}$ and $B^{(k)}$ are learnable weight matrices, and σ is a non-linear activation function. This process enables GNNs to capture both local and global relationships within the graph. However, GNNs may face challenges with deep graphs, leading to over-smoothing, or when distinguishing neighbors with varying degrees of importance [37].

Graph Attention Convolution (GATConv) is an advanced variant of GNNs that incorporates an attention mechanism to dynamically assign weights to neighboring nodes, rather than aggregating information uniformly. This allows the model to focus on more relevant neighbors, enhancing its ability to represent complex graph structures [38]. In GATConv, the attention coefficient $\alpha_{uv}$ between nodes $u$, $v$ is computed as follows:

$$e_{vu} = LeakyReLU\left(a^T[Wh_v \,||\, Wh_u]\right)$$

$$\alpha_{vu} = \frac{exp(e_{vu})}{\sum_{u' \in N(v)} exp(vu')}$$

Here, $h_v$, $h_u$ are the feature vectors of nodes $v$ and $u$, $W$ is a linear transformation matrix, $a$ is a learnable attention vector, $||$ denotes vector concatenation, and LeakyReLU is the activation function. The updated feature vector of node $v$ is computed as:

$$h_v' = \sigma\left(\sum_{u \in N(u) \cup \{v\}} \alpha_{vu} W h_u\right)$$

To improve stability, GATConv often employs multi-head attention, where multiple sets of attention coefficients are computed in parallel and combined. GATConv outperforms traditional GNNs by adapting to heterogeneous graph structures, mitigating over-smoothing, and enhancing performance in graph-related tasks.

### 2.4. Molecular Docking

Molecular Docking is a method used to study the binding affinity between active compounds and biological targets. The 3D structures of molecules will be downloaded from the PubChem database in SDF format and converted to MOL format using Open Babel software. For molecules not available in the database, their structures will be constructed and converted to the appropriate format. Biological targets will be retrieved from the Protein Data Bank, processed, and subjected to molecular docking studies using AutoDock 4.2.6 software. Finally, for each target-small molecule pair, the most spatially suitable conformations will be selected and their interaction results will be visualized using BIOVIA Discovery Studio Visualizer.

## III. Results and discussion

### 3.1. Network Pharmacology

#### 3.1.1. Targets prediction of active compounds of Garcinia cowa

Based on the collection of results from scientific publications combined with experimentally isolated active compounds, 87 compounds were obtained. All compounds were converted into SMILES format and their targets were predicted using the SwissTargetPrediction tool. Subsequently, the main active compounds were screened by eliminating those with fewer targets at a probability threshold above 0.1, resulting in seven key active compounds: Rubraxanthone, Garcinone D, Norcowanin, Cowanol, and Cowaxanthone. These compounds exhibited the highest degree of connectivity, corresponding to the largest number of targets above the 0.1 probability threshold.

### 3.1.2. Target-target interactions network of Garcinia cowa related to Hela cancer cell

The results collected from the GeneCard database using the keyword "HeLa" included 20,064 targets. After filtering and constructing a Venn diagram, 62 targets were identified as the common intersection between targets related to HeLa cancer cells and targets associated with the main active compounds from *Garcinia cowa*. To determine the primary targets, a protein-protein interaction (PPI) network was established using the STRING database, revealing TNF (PDB ID: 7ASY), CTNNB1 (PDB ID: 6M93), SRC (PDB ID: 7A3D), NFKB1 (PDB ID: 1MDK), and MTOR (PDB ID: 7PE7) as key central targets in the network. These are hypothesized to be the primary potential targets underlying the mechanism of cytotoxic activity against HeLa cells by the active compounds in *Garcinia cowa*.

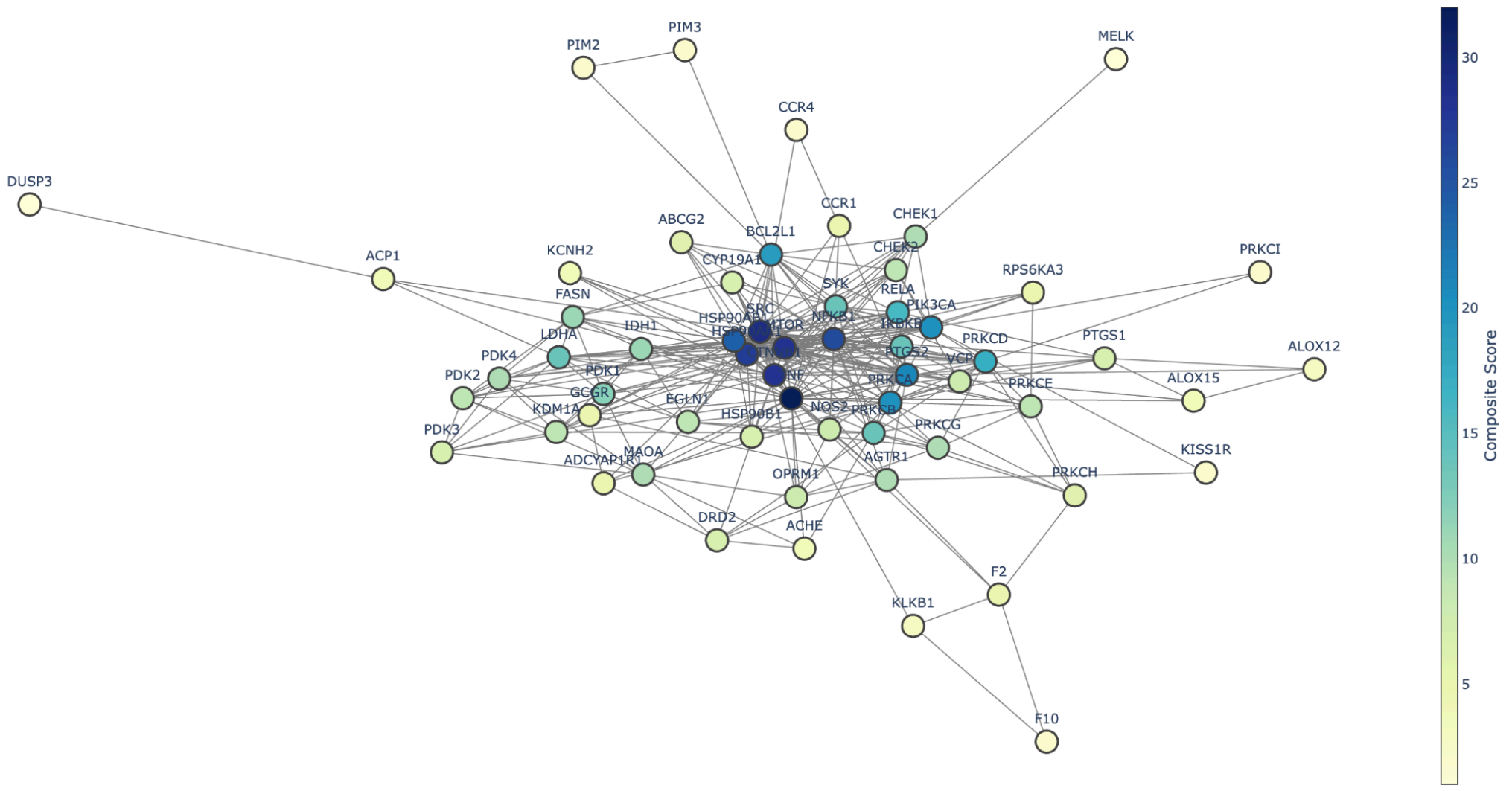

Figure 2: Protein-protein interaction networks and composite score

### 3.1.5. GO and KEGG analysis

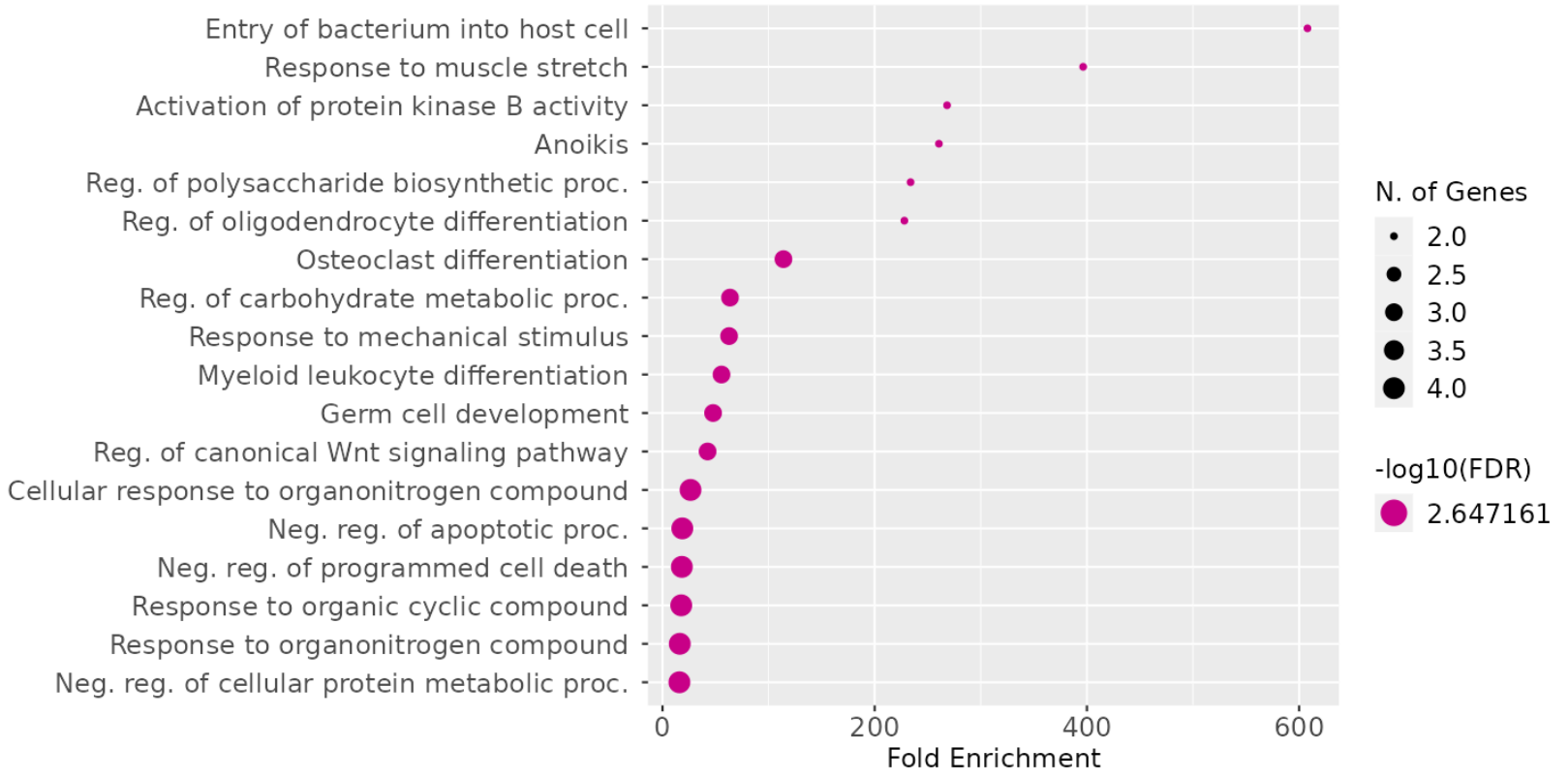

Figure 3: The result of Gene Ontology - Biological Process analysis

The GO-BP analysis reveals significant enrichment of the gene set in several key biological processes. Notably, processes related to cellular responses to organic compounds, such as "Cellular response to organonitrogen compound" and "Response to organic cyclic compound," are prominent, indicating that these genes are heavily involved in sensing and responding to organic compounds, potentially linked to detoxification mechanisms or intracellular signaling regulation. Additionally, processes like "Negative regulation of programmed cell death" and "Negative regulation of apoptotic process" are strongly enriched, suggesting a role for these genes in promoting cell survival under stress conditions. Furthermore, enrichment in processes such as "Myeloid leukocyte differentiation," "Response to mechanical stimulus," and "Regulation of canonical Wnt signaling pathway" indicates that the gene set may be involved in immune cell development, responses to mechanical stimuli, and regulation of critical signaling pathways in development and tissue regeneration. Processes like "Osteoclast differentiation" and "Regulation of carbohydrate metabolic process," while less enriched, still show evidence of involvement in metabolic balance and bone tissue remodeling.

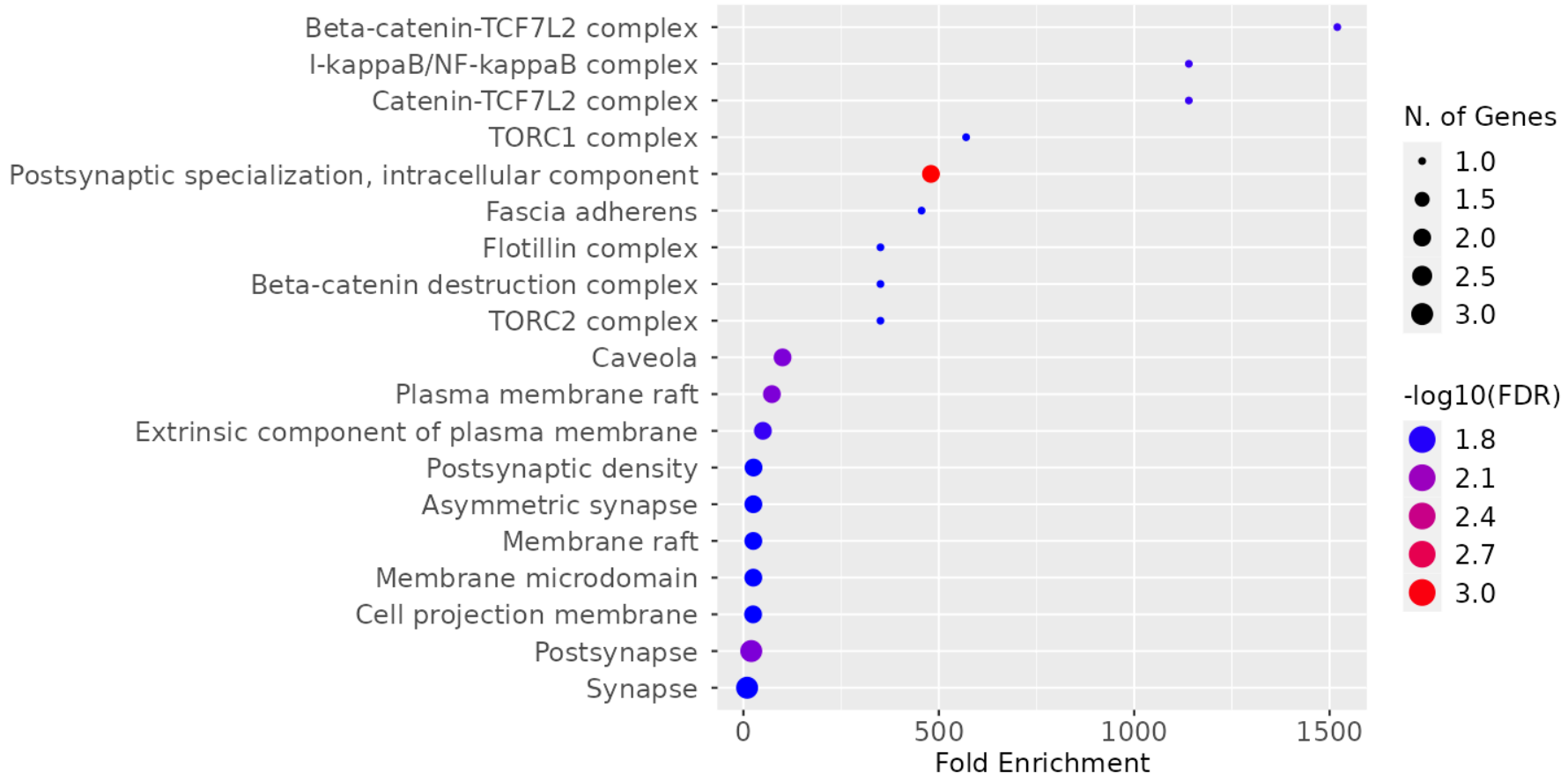

Figure 4: The result of Gene Ontology - Cellular Component analysis

The GO-CC analysis reveals significant enrichment of the analyzed gene set in various specialized structures related to the cell membrane and neural signaling regions. Specifically, GO clusters such as "Synapse," "Postsynapse," and "Postsynaptic density" stand out with high enrichment, indicating that these genes play a critical role in the formation and maintenance of synaptic structures, particularly in the postsynaptic region, where signals from presynaptic neurons are received. This suggests that the gene set is closely associated with processes regulating neural signal transmission or synaptic plasticity (the adaptability of synapses). Additionally, membrane-related components like "Plasma membrane raft," "Membrane raft," and "Membrane microdomain" are significantly enriched. These microdomains often serve as platforms for organizing signaling proteins/machinery, facilitating efficient intracellular signal transduction, and are also linked to membrane stability and the sorting of extracellular signals. Clusters related to protein complexes, such as "Beta-catenin-TCF7L2 complex," "Catenin-TCF7L2 complex," "I-kappaB/NF-kappaB complex," and "Beta-catenin destruction complex," indicate strong involvement of these genes in regulating Wnt/β-catenin and NF-κB signaling pathways. These are critical signaling pathways for controlling cell growth, division, and inflammatory responses, suggesting potential roles in cancer, tissue development, and immune responses. Notably, clusters like "TORC1 complex" and "TORC2 complex" also appear in the results, pointing to the genes' roles in regulating cellular metabolism, growth, and stress responses through the mTOR pathway a key signaling axis in cellular physiology and pathologies such as cancer and metabolic disorders. Furthermore, enrichment in structures like "Caveola" and "Fascia adherens" indicates involvement in mechanical membrane structures, supporting the sensing of mechanical forces and regulation of mechanosignaling, as well as cell adhesion.

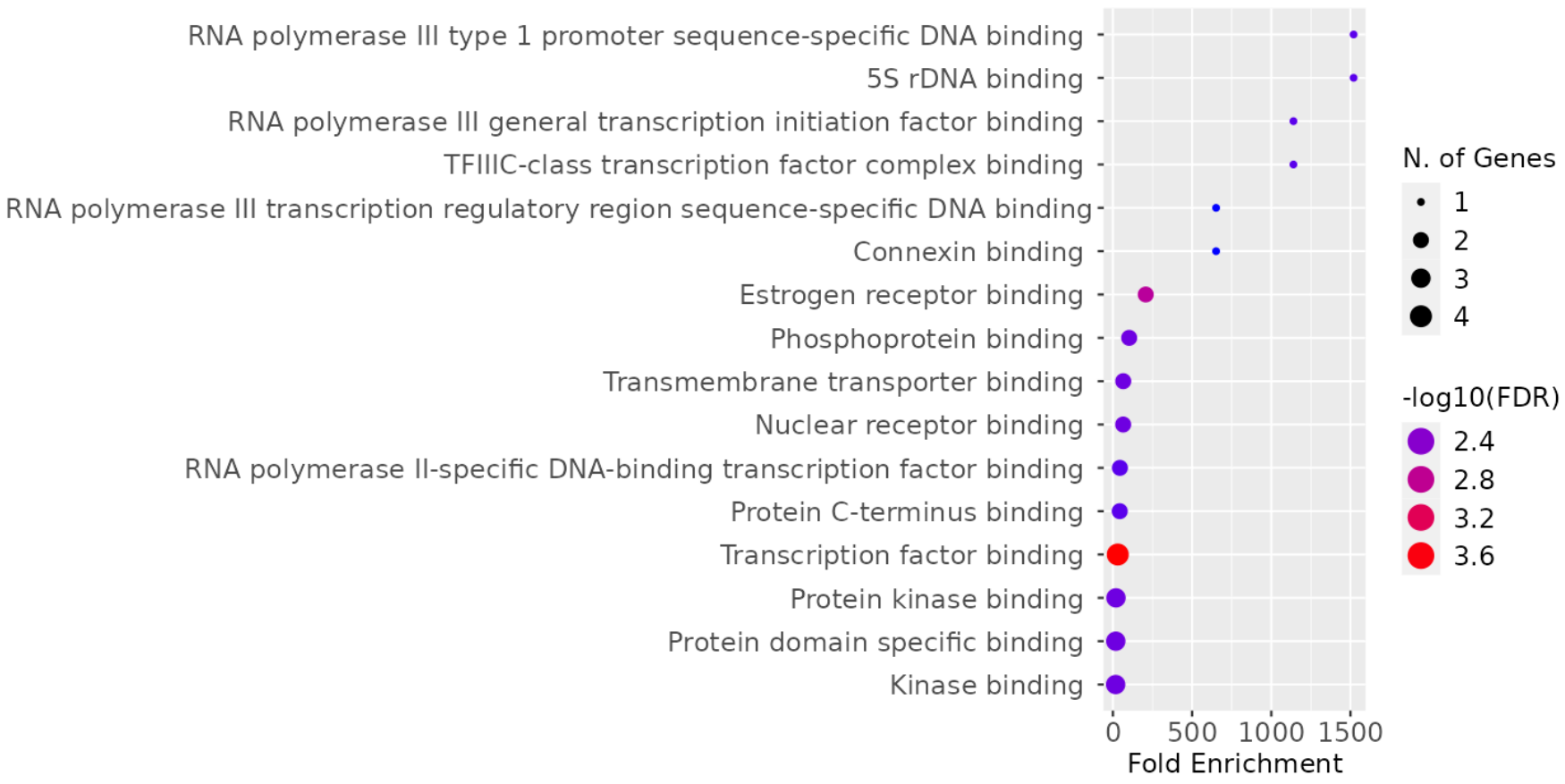

Figure 5: The result of GO - Molecular Function analysis

Based on Figure 5, it is evident that certain functions such as "Transcription factor binding", "Protein kinase binding", and "Estrogen receptor binding" not only exhibit relatively high fold enrichment, but are also associated with a larger number of genes and high -log10(FDR) values. This indicates strong statistical significance and suggests that these functions are likely to play important roles in the analyzed gene set. In contrast, functions like "RNA polymerase III type 1 promoter sequence-specific DNA binding" and "5S rDNA binding" show very high fold enrichment but are linked to only 1–2 genes and have high FDR values (indicated by blue color), making them statistically less reliable.

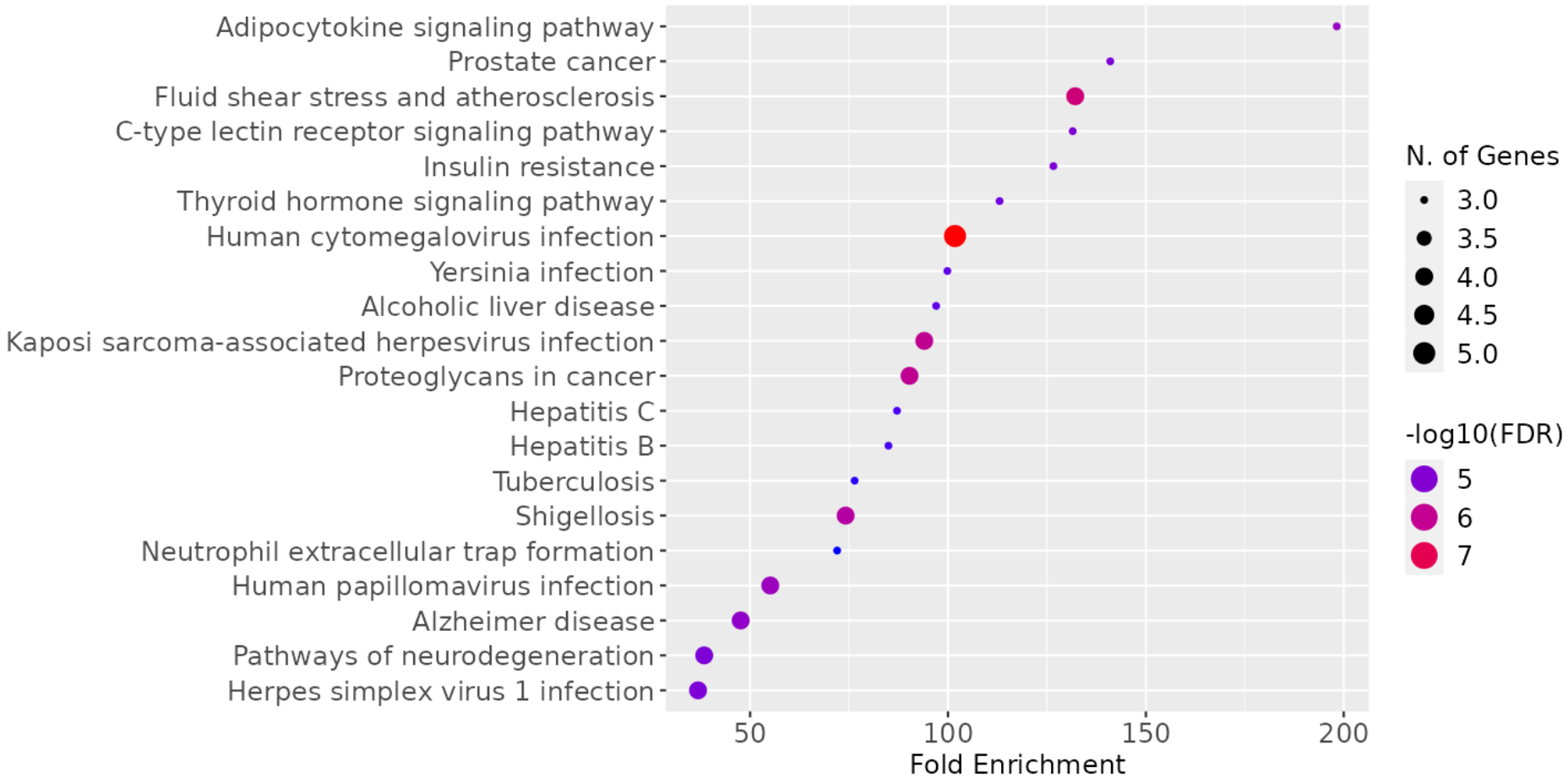

Figure 6: The result of KEGG pathway analysis

The KEGG pathway analysis reveals numerous pathways associated with inflammation, cancer, viral infections, and neurodegenerative processes. Several prominent pathways, such as “Human cytomegalovirus infection”, “Kaposi sarcoma-associated herpesvirus infection”, and “Yersinia infection”, show high fold enrichment, involve multiple genes, and have notably high -log10(FDR) values (indicated in red), suggesting strong statistical significance. This indicates that responses related to viral and bacterial infections may play a key role in the biological context of the sample. Additionally, pathways associated with cancer (e.g., “Prostate cancer”, “Proteoglycans in cancer”) and innate immunity (e.g., “Neutrophil extracellular trap formation”) are also significantly enriched, suggesting active inflammatory or immune responses. Some metabolic pathways, such as “Insulin resistance” and “Adipocytokine signaling pathway”, also appear with very high fold enrichment; however, they involve fewer genes and show lower statistical significance (indicated in purple), thus requiring more cautious interpretation. Overall, the results suggest that immune responses, viral infections, and cancer-related processes may represent key biological mechanisms in the analyzed gene set.

### 3.2. Cytotoxicity prediction using Machine learning

| | Decision Tree | Random Forest | XGBoost | GNN |
|---|---|---|---|---|
| **R2** | 0.35 | 0.76 | 0.81 | 0.88 |
| **RMSE** | 0.24 | 0.14 | 0.13 | 0.09 |

Table 1: Results of the Machine Learning model before and after using data augmentation

| Compounds | pIC50 true values | pIC50 predicted values |
|---|---|---|
| Rubraxanthone | 5.11 | 5.14 |
| Garcinone D | 4.92 | 4.89 |
| Norcowanin | 4.87 | 4.81 |
| Cowanol | 4.91 | 4.92 |
| Cowaxanthone | 4.80 | 4.78 |

Table 2: Comparison results of actual values and predicted values of the GNN model

The combined results from Table 1 and Table 2 reveal a significant difference in model performance before and after applying data augmentation techniques. Prior to augmentation, the Decision Tree model exhibited the poorest performance, with an $R^2$ of only 0.35 and an RMSE of 0.24. In contrast, the Graph Neural Network (GNN) stood out with an $R^2$ of 0.88 and an RMSE of 0.09, demonstrating its superior ability to extract structural information from SMILES data. After data augmentation, the performance of all models improved markedly. Notably, the Decision Tree, which previously underperformed, saw a dramatic increase in $R^2$ to 0.87 and a reduction in RMSE to 0.11, indicating that augmentation enabled the model to learn more complex separations in the feature space. Stronger models like XGBoost and GNN further solidified their leading positions, achieving $R^2$ values of 0.97 and 0.98, respectively, with very low RMSEs (0.04 and 0.02), reflecting their capability to model non-linear relationships between features and the target pIC50 values. This suggests that GNN not only benefits from augmented data but also effectively generalizes information for compounds within the same structural class. However, it is crucial to note that the training dataset was carefully filtered to retain only bioactive compounds with a primary xanthone scaffold, meaning the current models are optimized for molecules with high structural similarity to the original data. To broaden the predictive scope, it will be necessary to redesign the data collection, preprocessing, and augmentation pipeline to ensure greater structural diversity in the training set. Nevertheless, to mitigate model complexity, training could be conducted on specific compound groups, allowing the model to focus on predicting the activity of a single group. Additionally, data augmentation plays a pivotal role in enhancing model performance. By generating simulated samples while preserving core chemical features, this technique enables models to capture latent variations in molecular data, thereby improving generalization. However, a prerequisite is that the original dataset must be sufficiently accurate and large; otherwise, augmentation may amplify biases or lead to subtle overfitting with unwanted noise. Furthermore, for graph-based machine learning models, extracting detailed features for nodes and edges is essential to ensure optimal performance.

**3.3. Molecular docking**

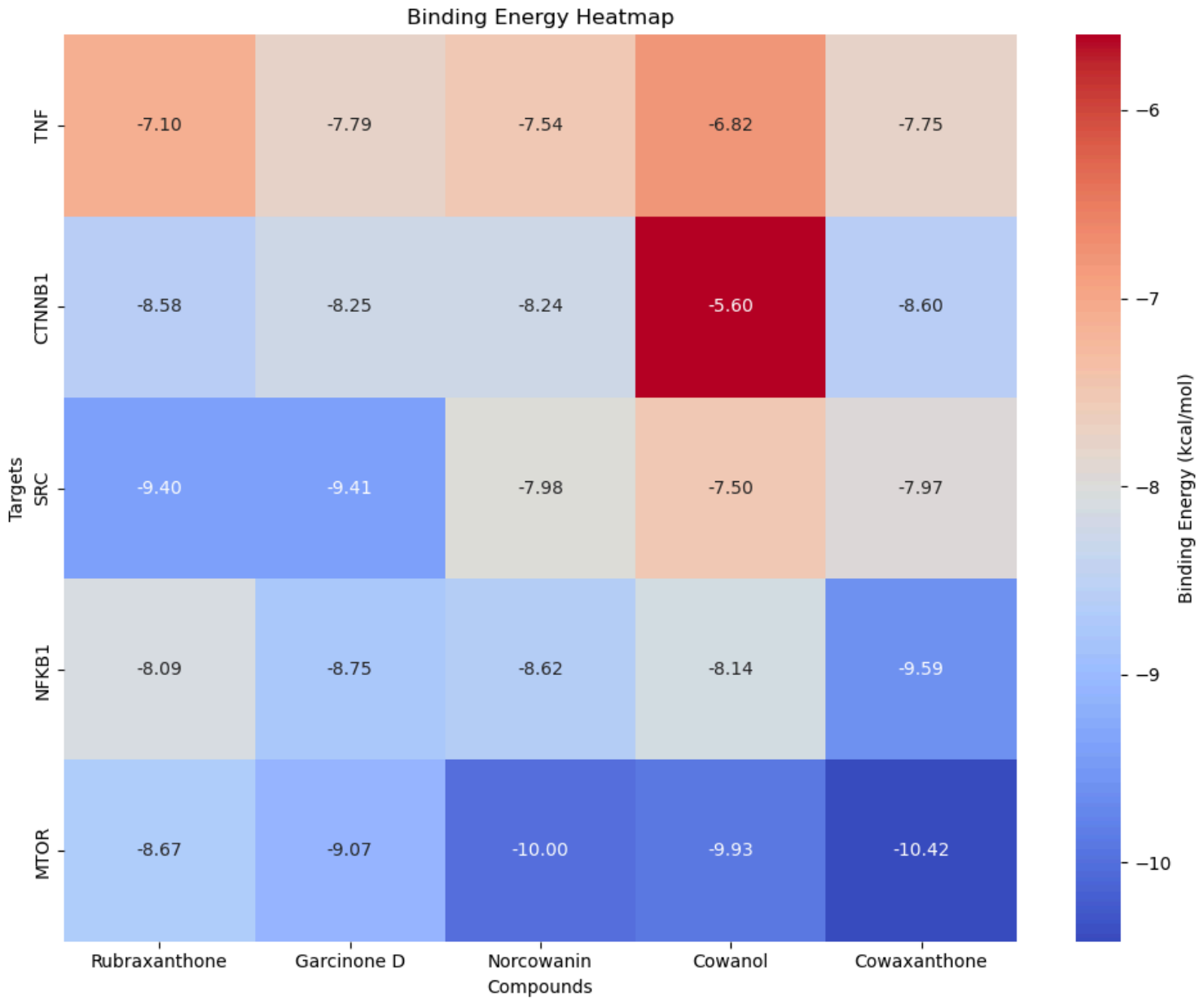

Figure 7. Heatmap of binding energies between compounds and targets

Based on Figure 7, most compounds exhibit high binding affinity to the targets, with binding energies ranging from -6.82 to -10.42 kcal/mol. However, Cowanin does not show effective interaction with CTNNB1, as its binding energy is only -5.60 kcal/mol. With binding energies between -6.82 and -7.97 kcal/mol considered good and values lower than this indicating strong interactions, these results highlight compounds like Cowaxanthone and Norcowanin, particularly with MTOR (-10.42 and -10.00 kcal/mol), as potential targets for the cytotoxic activity on HeLa cells of Garcinia cowa. These molecular docking results further validate the findings of previous network pharmacology studies by suggesting potential targets for the bioactive compounds in this medicinal plant. Moreover, they pave the way for promising future research by integrating various investigative approaches.

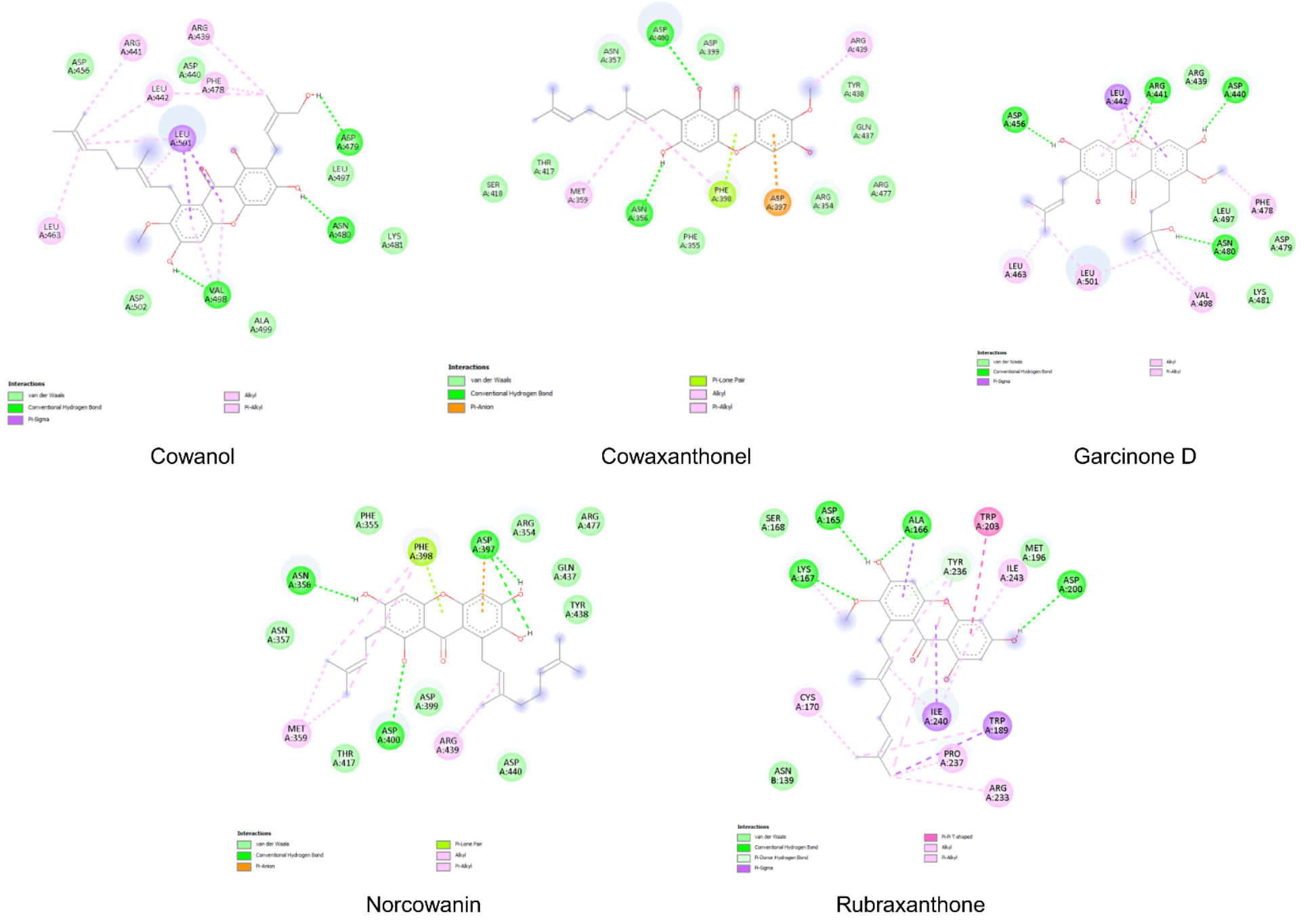

Figure 8. Interactions of Xanthone Compounds from Garcinia cowa with Protein Targets CTNNB1

The molecules all demonstrate the ability to interact with the target protein CTNNB1, as shown in Figure 8, through various types of non-covalent interactions such as hydrogen bonds, van der Waals forces, and hydrophobic interactions (pi-alkyl, alkyl). A common feature among the compounds is the presence of an aromatic ring core, which facilitates pi interactions with amino acids like ARG, LEU, PHE, or TYR. However, each compound exhibits distinct characteristics in terms of interaction types and binding positions. Rubraxanthone stands out with the most diverse and complex interaction profile, including pi-pi T-shaped, pi-sigma, and interactions with amino acids such as TRP (TRP203, TRP189) and ARG, indicating a strong binding affinity to the active site. Notably, the hypothesis suggests that the pi-sigma interactions with ILE240 and TRP198, along with the pi-pi T-shaped interaction with TRP203, contribute to the molecule's excellent binding energy. In contrast, Garcinone D has fewer interactions, primarily van der Waals and hydrogen bonds, but still demonstrates good binding affinity with a binding energy of 8.25 kcal/mol. Cowaxanthone exhibits additional pi-anion interactions, similar to Norcowanin, with the amino acid PHE398, suggesting that pi-anion interactions with this residue may enhance binding affinity. Meanwhile, Cowanol interacts with central regions of the active site and, despite engaging with several amino acid residues, does not exhibit strong binding affinity due to its very low binding energy. Nevertheless, the results highlight the diversity in interaction types and binding positions, indicating the varying potential of each compound in modulating or inhibiting CTNNB1 activity.

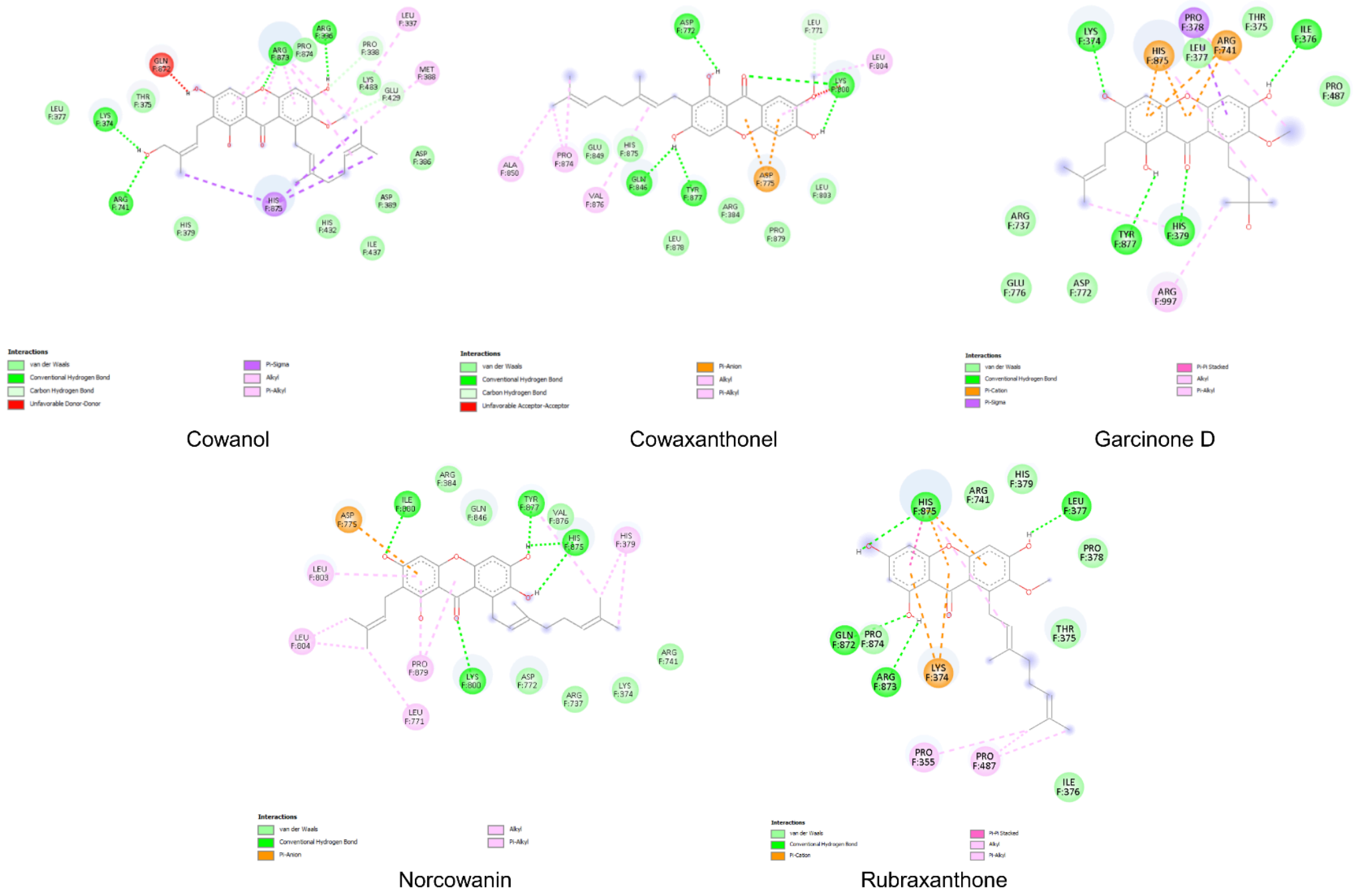

Figure 9. Interactions of Xanthone Compounds from Garcinia cowa with Protein Targets MTOR

For the target MTOR, all active compounds demonstrate good binding affinity to the target. Based on Figure 9, HIS875 is a common amino acid across all compounds, with various interaction types (Cowanol: Pi-Sigma, Cowaxanthone: Carbon Hydrogen Bond, Norcowanin: Conventional Hydrogen Bond, Garcinone D: Pi-Cation, and Rubraxanthone: Conventional Hydrogen Bond). Additionally, the Pi-Anion interaction between the drug molecule and ASP775 is shown to strongly contribute to the binding affinity, particularly for Norcowanin and Cowaxanthone, which exhibit the highest binding energies of -10.00 and -10.42 kcal/mol, respectively. Furthermore, the alkyl interaction with HIS379, observed in Norcowanin, suggests better interaction potential compared to the van der Waals interactions of Cowanol and the Conventional Hydrogen Bond of Garcinone D. This could be help for the future optimization of drug molecules targeting MTOR.

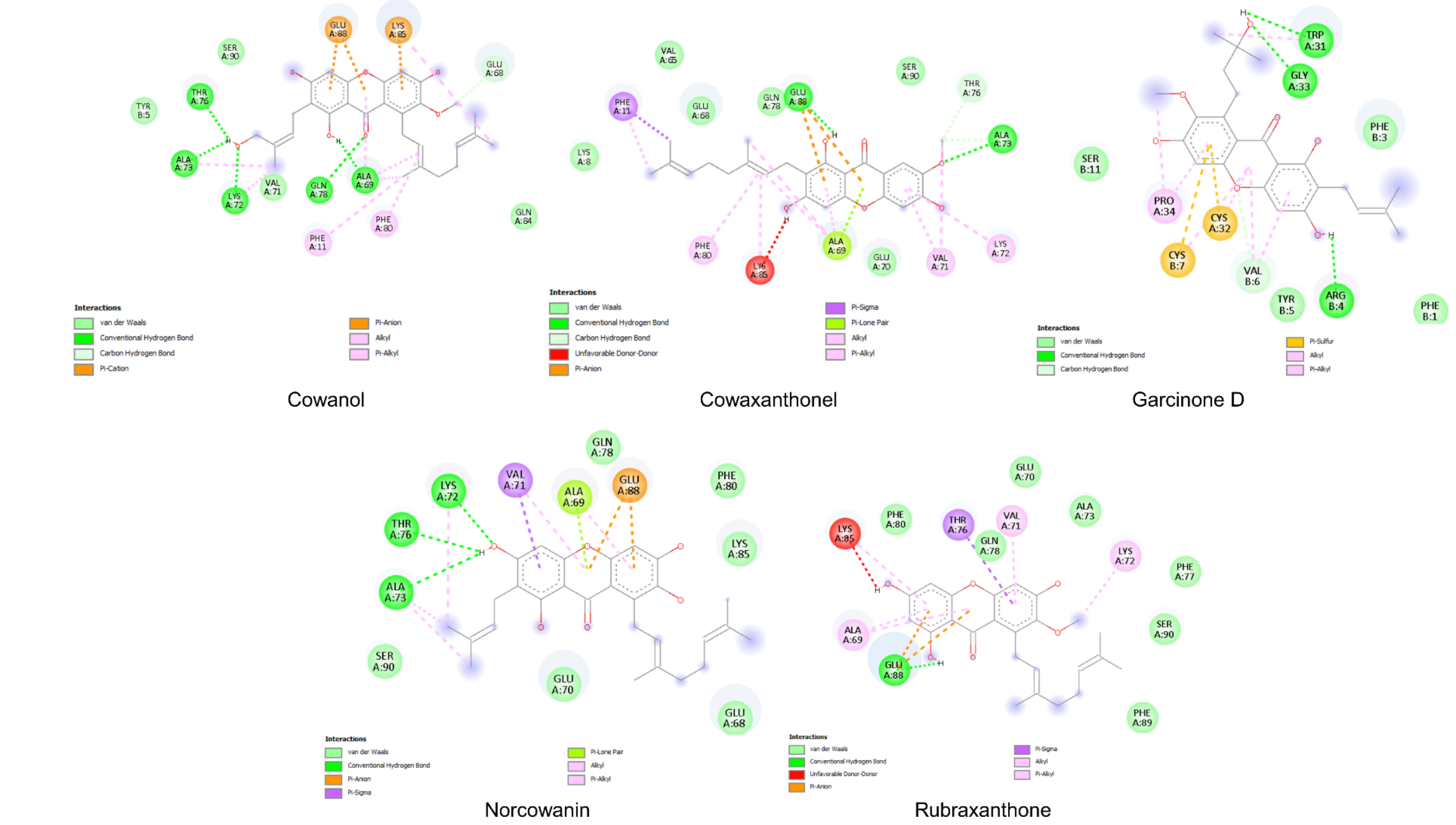

Figure 10. Interactions of Xanthone Compounds from Garcinia cowa with Protein Targets  NFKB1

Although there is a variety of interactions with the target, the binding energies of the compounds are not higher than those with the MTOR target. Even so, this is still a target where the active compounds show very good interactions, with binding energies ranging from -8.09 to -9.59 kcal/mol. Except for Garcinone D, the other compounds interact with similar amino acids, such as GLU88, LYS85, PHE80, SER90, THR76, and so on. Meanwhile, Garcinone D has the second-highest binding affinity with the target, suggesting a unique interaction mechanism for this compound.

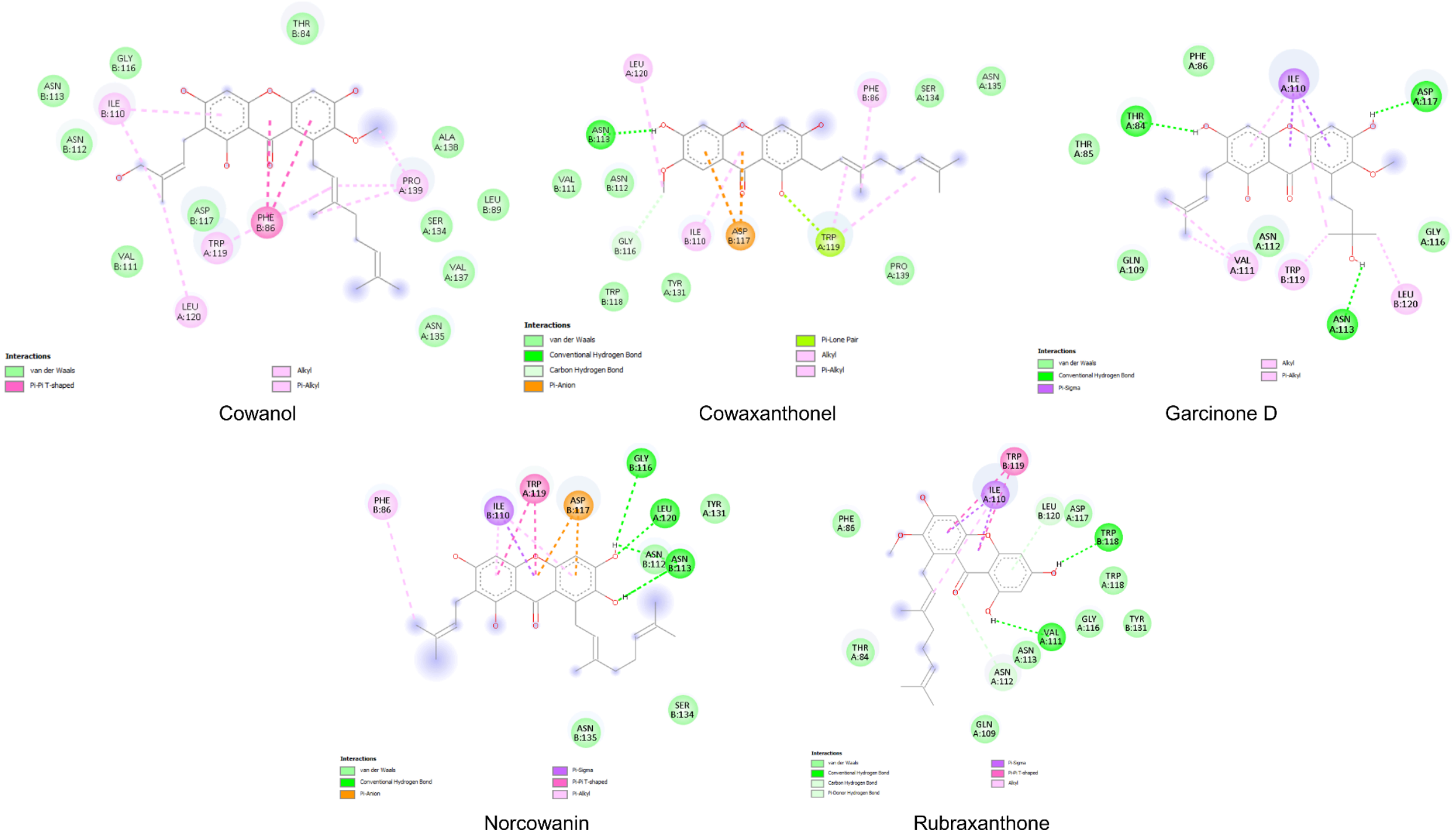

Figure 11. Interactions of Xanthone Compounds from Garcinia cowa with Protein Targets SRC

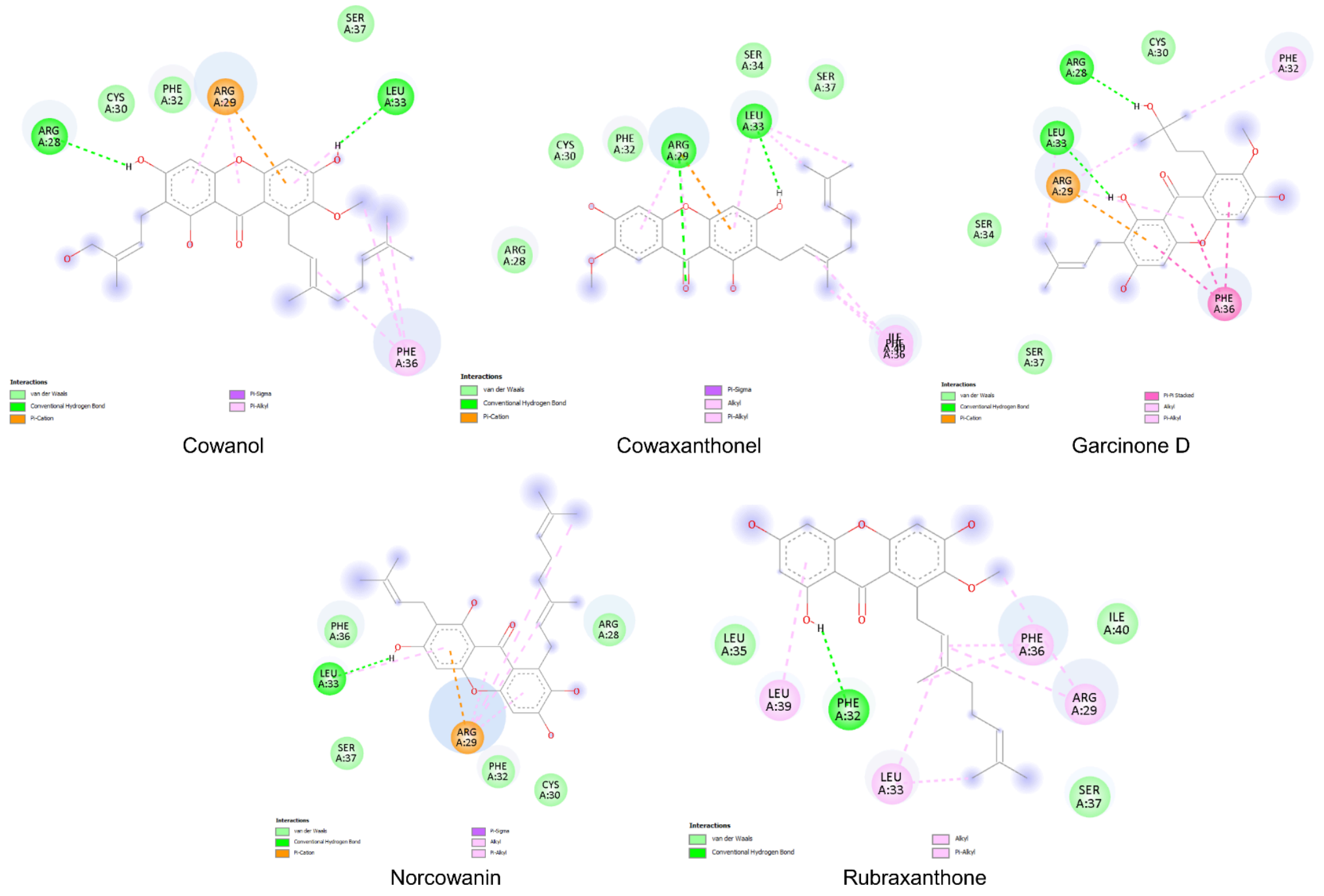

Figure 12. Interactions of Xanthone Compounds from Garcinia cowa with Protein Targets TNF

For targets like TNF and SRC, most compounds show good binding interactions. However, Rubraxanthone and Garcinone D exhibit very strong interactions with the SRC target, with binding energies of -9.40 and -9.41 kcal/mol, respectively. This may be due to the Pi-Sigma interaction with the amino acid ILE110. Additionally, Norcowanin and Cowaxanthone also show Pi-Anion interactions with ASP117, indicating a higher interaction potential compared to Cowanol. In contrast, these two compounds have lower binding energies compared to Garcinone D and Rubraxanthone because their interactions with ASP117 are Conventional Hydrogen Bonds and van der Waals forces. Similarly, with PHE86, van der Waals interactions provide better interaction potential than Pi-Alkyl interactions. For the TNF target, although it is classified as having good binding energy, it shows the weakest interactions compared to the other four targets. This might be due to the diversity of interactions as well as the suitability of the three-dimensional structure.

**IV. Conclusion**

This study has demonstrated the robust integration of advanced computational methodologies, including network pharmacology, machine learning, and molecular docking, to elucidate the cytotoxic potential of bioactive compounds from Garcinia cowa against HeLa cancer cells. The network pharmacology analysis successfully identified key active compounds Rubraxanthone, Garcinone D, Norcowanin, Cowanol, and Cowaxanthone and their primary protein targets (TNF, CTNNB1, SRC, NFKB1, and MTOR), providing critical insights into the molecular mechanisms underlying their anti-cancer activity. The application of machine learning, particularly the Graph Neural Network model, showcased exceptional predictive performance with an $R^2$ of 0.98 and an RMSE of 0.02 post-data augmentation, highlighting its capability to accurately predict pIC50 values for compounds with xanthone scaffolds. These results underscore the power of graph-based deep learning in capturing complex molecular structural relationships, offering a scalable approach for quantitative structure-activity relationship modeling in drug discovery. Molecular docking further validated these findings by revealing strong binding affinities, notably for Cowaxanthone and Norcowanin with MTOR (-10.42 and -10.00 kcal/mol, respectively), reinforcing the potential of these compounds as lead candidates. Collectively, this integrative in silico framework not only accelerates the identification and optimization of novel anti-cancer agents but also sets a precedent for leveraging machine learning-driven approaches to navigate the complexities of natural product-based drug development. Future efforts should focus on expanding the structural diversity of training datasets and incorporating experimental validation to translate these computational insights into clinical applications, ultimately advancing precision oncology.